\documentclass[12pt]{article}

\usepackage{graphicx}

\title{Long Term and Short Term Effects of Perturbations in a Immune Network Model}

\author{Rita Maria Zorzenon dos Santos \\ 
Mauro Copelli \\
\small Laborat\'orio de F\'{\i}sica Te\'orica e Computacional \\ 
\small Departamento de F\'\i sica, Universidade Federal de Pernambuco \\
\small Cidade Universit\'aria, 50670-901, Recife, PE, Brazil} 
\date{\today}

\begin{document}

\maketitle

\abstract{In this paper we review the trajectory of a model proposed
by Stauffer and Weisbuch in 1992 to describe the evolution of the
immune repertoire and present new results about its dynamical
behavior.  Ten years later this model, which is based on the ideas of
the immune network as proposed by Jerne, has been able to describe a
multi-connected network and could be used to reproduce immunization
and aging experiments performed with mice. Besides its biological
implications, the physical aspects of the complex dynamics of this
network is very interesting {\it per se\/}. The immunization protocol
is simulated by introducing small and large perturbations (damages),
and in this work we discuss the role of both. In a very recent paper
we studied the aging effects by using auto-correlation functions, and
the results obtained apparently indicated that the small perturbations
would be more important than the large ones, since their cumulative
effects may change the attractor of the dynamics. However our new
results indicate that both types of perturbations are important. It is
the cooperative effects between both that lead to the complex behavior
which allows to reproduce experimental results.}

\section{Introduction}

The main task of the immune system is to protect the organism against
dangerous elements: {\it antigens} (virus, bacteria, poison, cell
residues, etc). Depending on the antigen the immune system may elicit
different kinds of responses: the cell-mediated immune response or the
humoral response. The models discussed in this paper are related to
the humoral responses generated by B cells (one of the main classes of
lymphocytes), which are the cells that produce the antibodies. The
antibodies produced by a given population of B cells are replicas of
its molecular receptor. Each molecular receptor exhibited by a given B
cell population recognize different recognition sites (epitopes) of
the antigen by lock-key interactions.

The clonal selection theory~\cite{janeway} is the most accepted theory
about the B cells and was proposed by Burnet in $1959$. It states that
the antigen chooses by pattern recognition the clones of B cells
(population of B cell and antibodies) that will proliferate. In order
to recognize any foreign (or dangerous) element the immune repertoire
must be complete. According to estimates the human immune system is
able to express at least the order of $10^6$ different
receptors~\cite{janeway}. Due to the completeness of the repertoire,
the immune system would be able to recognize and be recognized
(recognizing epitopes of its own antibodies), therefore the same
mechanism of recognition should work for both antibody-antigen and
antibody-antibody reactions. In 1974 Jerne~\cite{Jerne}, taking into
account these different mechanisms, suggested that when the antigen is
presented to the organism it will activate a set of clones of B cells,
leading to the production of specific antibodies; those antibodies
will in turn be recognized by a second set of clones activating them,
and so on. Due to the interplay of the mechanisms of activation and
suppression, this chain of reactions will be finite, preventing the
``information'' from ``percolating'' through the entire
system. Therefore the immune system functions as a functional
network~\cite{Jerne}, with complexity comparable to the nervous
system. Since its proposal, only few evidences support the existence
of the immune network theory~\cite{coutinho89, holmberg}, and those
evidences suggest that if the network exists only $20\%$ of the
lymphocytes will be activated, while the rest of the clones will form
a pool of immunocompetent lymphocytes that are able to recognize any
antigen.

In what follows we define in section~\ref{model} a model introduced by
Stauffer and Weisbuch in 1992~\cite{sw}, which incorporates the main
concepts of Jerne's immune network theory. In section~\ref{previous
results} we review the main results obtained in the literature,
focusing on both the physical~\cite{zsb95,bzs97} and biological
aspects~\cite{zsb98}, since among many other attempts to model what an
immune network could be, this was the first that could reproduce the
behavior of a real immune system.  In section~\ref{results} we present
new results about the short and long time behaviors of the
multi-connected network under small and large perturbations. We also
discuss the results obtained about the aging effects in a very recent
work~\cite{Copelli03} on the light of these new results.  In our
concluding remarks (section~\ref{conclusao}) we address the future
prospects from the physical and biological points of view.

\section{The Model}
\label{model}

In 1992 Stauffer and Weisbuch (SW)~\cite{sw} introduced a cellular
automata model to describe the evolution of the B cell
repertoire. This model was inspired in a previous one (based on a
difference equation approach) introduced by de Boer, Segel and
Perelson (BSP)~\cite{bsp} using a shape-space formalism to describe a
large-scale immune network. Using a discrete shape-space formalism SW
associate each point of a $d$-dimensional space to a different
molecular receptor (or a clone of B cell). In this way each receptor
is characterized by $d$ properties, and its neighbors in the shape
space will correspond to molecular receptors that differ from it by
one of these properties (according to the estimates based on a
continuous approach~\cite{peroster}, if the notion of shape space is
relevant, then $d\geq 5$). 

To each receptor they associate a three-state cellular automaton
$B(\vec r,t)$ that will describe the concentration of the population
characterized by this receptor $\vec r$ over the time: low ($B(\vec
r,t)=0$), intermediate ($B(\vec r,t)=1$) and high ($B(\vec r,t)=2$).
Following the ideas of the immune network theory, the interactions
among different populations are based on complementarity (lock-key
interaction) and are defined by deterministic rules. The time
evolution of the cellular automaton is based in a deterministic
non-local rule: population $B(\vec r,t)$ at site $\vec r$ is
influenced by the populations at site $-\vec r$ (its mirror image or
complementary shape) and its nearest-neighbors ($-\vec r + \delta \vec
r$) (representing defective lock-key interactions). The influence on
the population at site $\vec r$ caused by its complementary
populations is described by the field $h(\vec r,t)$:

\begin{equation}
h(\vec r,t)=\sum_{ \vec r^{\,\prime}\in (-\vec r + \delta \vec r)}
B( \vec{r}^{\, \prime},t)
\end{equation}
where for a given $\vec r$ the sum runs over the complementary shape $
\vec r^{\, \prime} = -{\vec r}$ and its nearest neighbors. Due to the
finite number of states of the population $B$, the maximum value of
the field $h( \vec r)$ is $h_{max}=2(2d+1)$. The rules are based on a
window of activation for each population which is inspired in a log
bell shaped proliferation function associated to the receptor
cross-linking involved in the B cell activation~\cite{zsb95,sw,bsp}.
There is a minimum field necessary to activate the proliferation of
the receptor populations ($\theta_1$), but for high doses of
activation (greater than $\theta_2$) the proliferation is
suppressed. The updating rule may be summarized as:

\begin{equation} 
\label{eq:dyn}
B( \vec r,t+1)= \left\{ \begin{array}{ll}
B( \vec r,t) +1 \qquad & {\rm if} \qquad \theta_1 \leq h( \vec r,t) \leq \theta_2 \\
B(\vec r,t) - 1 &{\rm otherwise} \end{array} \right.
\end{equation}
but no change is made if it would lead to $B=-1$ or $B=3$.  We define
the densities of sites in state $i$ at time $t$ as
$B_i(t)$ ($i=1$, 2, 3).

The initial configurations are randomly generated according to the
parameter control $x$, which determines the initial concentrations:
$B_1(0)= B_2(0)=x/2$, while the remaining $L^d(1-x)$ sites are
initiated with $B(\vec r,0)=0$.

\section{Results obtained from previous studies}
\label{previous results}

The above definition of the model corresponds already to the modified
version which has been proposed in Ref.~\cite{zsb95}. SW have shown
that for the original version of the model~\cite{sw}, there is a
stable-chaotic transition on the behavior of the discrete system for
$d \ge 4$, when varying the parameter $x$. As pointed out by the
authors, according to the characteristics of those behaviors, none of
them would be appropriate to describe the evolution of the immune
repertoire.  

One of us has studied how the system attains the chaotic regime, by
studying the transient times and periods of the original model close
to the transition~\cite{zs93}.  It was observed that close to the
transition, the system is trapped in cycles with large but finite
periods.  In 1995 Zorzenon dos Santos and Bernardes~\cite{zsb95}
introduced the modified version of SW model described above and have
extensively studied the behavior of this model in the parameter space.
Differently from SW, they obtained~\cite{zsb95} a stable-chaotic
transition for $d \ge 2$, and mapped all the behaviors on ``phase''
diagrams for different set of parameters, showing that there was a
broad transition region between stable and chaotic behaviors.

In the following work~\cite{bzs97} Bernardes and Zorzenon dos Santos
have investigated the dynamical behavior of the model on this broad
transition region. They have observed an aggregation-disaggregation
dynamics, with clusters splitting and fusing along the time, as a
multi-connected network.  The authors have also studied the behavior
of the system in the transition region when subjected to antigen
presentation. The antigen presentation was simulated by flipping
randomly chosen populations from the inactive state to the highly
activated state, reproducing in this way the activation of the
populations caused by the presence of an antigen. The results obtained
by the standard spread of damage procedure~\cite{jancar94} indicate
that after the antigen presentation, the perturbation first increases
and then relaxes after a few time steps, indicating that some
information about this perturbation is incorporated by the
system. These results suggested that this model would be a good
candidate to model a real immune system behavior. Zorzenon dos Santos
and Bernardes then applied it~\cite{zsb98} to reproduce immunization
and aging experiments performed with mice under multiple antigen
presentations~\cite{bruno97,ana97,lahmann} obtained by the group of Nelson
Vaz.

In the experiment of immunization~\cite{bruno97} six mice of the same
linage are subjected to the following protocol: the antigen
presentations are produced by intra-venal injections of ova. The time
interval between the first and second presentation is 14 days, but
after that the time interval between consecutive presentations is
fixed to 7 days. Before each antigen presentation (except for the
first one) they measure the amount of specific antibodies in the
blood. Among many experiments reported by this group, Zorzenon dos
Santos and Bernardes chose the one showing a refractory (saturation)
behavior related to the immunization protocol~\cite{bruno97} that
could not be explained by clonal selection theory. However, the aging
experiment chosen~\cite{ana97,lahmann} is one among others obtained by the
group showing the same result: a reduction on the intensity of the
response as the system ages.

In order to simulate the immunization protocol~\cite{zsb98}, for each
sample they let the system evolve from its initial configuration to
$1000$ time steps, and associate this time step to the birth of the
simulated mouse. Then random small perturbations, in size and location
in the shape-space (using different time intervals between them), are
produced in order to simulate the noise to which the immune system is
subjected due to the environment, food ingestion, etc. By using the
arbitrary scale of 1 day corresponding to 5 time steps, it is possible
to simulate independent antigen presentations without over-exciting
the system~\cite{bzs97}. The immunization protocol is simulated by
large perturbations (one order of magnitude greater than the small
ones) being produced periodically every $7$ days. The protocol was
simulated for different samples ($10$) of young and old mice (8 and 24
weeks, respectively). The results could be interpreted on the light of
the immune network theory, the refractory behavior being associated to
the saturation of the multi-connected network in incorporating
information about the perturbation, and the aging effects related to
the loss of plasticity of the system.

Since there are only few evidences that support the existence of the
immune network, this model played a very important role in making the
connection between some experimental results that could not be
explained by clonal selection theory, and the immune network
theory. Moreover the theoretical results indicate that under multiple
antigen presentation the saturation is due not only to the number of
specific antibodies produced (as shown in the Elisa experiments) but
also the ability to incorporate information by including new
populations in the network. Since the number of populations belonging
to the network does not change significantly during the time
evolution, in order to add information, part of the information
already present in the network should be lost. Therefore the profile
of expressed antibodies would change after each antigen presentation,
reflecting such changes. This is an aspect that comes out of the
theoretical results and should be investigated by the experimentalists
when the appropriated tools become available. The techniques currently
available do not allow to identify the difference among the specific
antibodies. That would be a requirement to verify whether they change
after multiple presentations as suggested by the model. However, there
are some experimental evidences that it may happen~\cite{gerlinde}.

Once the complex multi-connected network generated by the model can be
used to reproduce the behavior of a real immune system, it becomes
interesting to investigate its dynamical properties from the physical
point of view. This investigation will allow to learn about the
dynamics of a (real) complex system operating out of equilibrium. The
characterization of such behavior started with the work of Bernardes
and Zorzenon dos Santos~\cite{bzs01}, who analyzed the distribution of
cluster sizes and the distribution of the permanence time (i.e. the
time interval during which each population remains activated or belong
to the multi-connected network). They found a characteristic cluster
size and a power law behavior for the distribution of permanence
times. The characteristic cluster size is associated to the loss of
plasticity while the power law distribution indicate that the
populations belonging to the network have no typical permanence time,
reflecting the fact the dynamical memory is generated by incorporating
information about the different antigens presented to the system.

\begin{figure}[!bt]
\begin{center}
\includegraphics[width=0.7\textwidth,angle=-90]{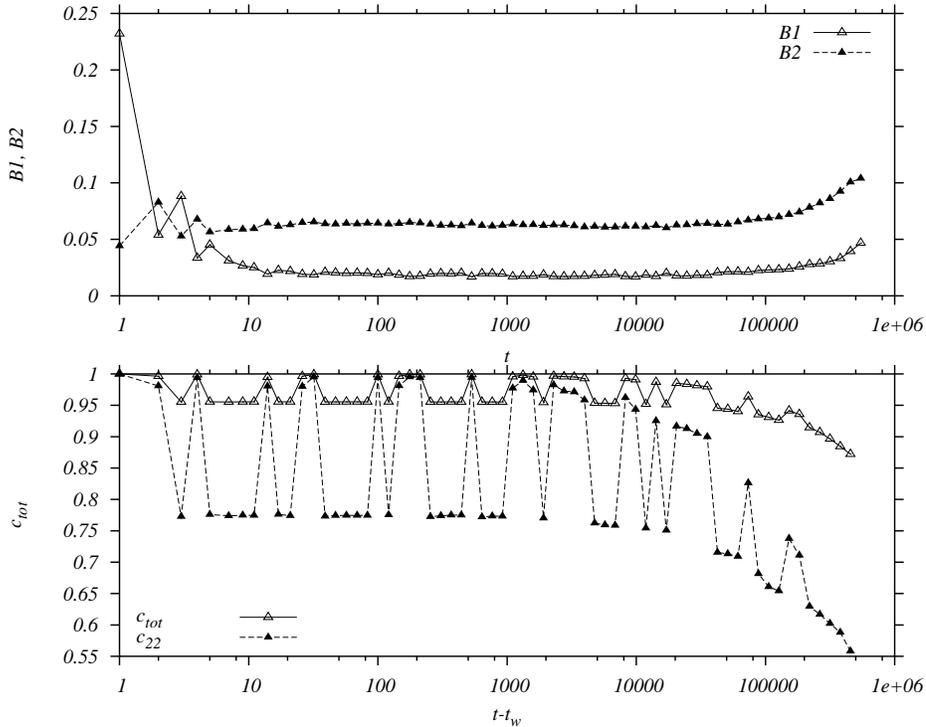}
\caption{\label{fig:correlation}Densities vs. time (upper panel) and
correlations vs. $t- t_w$ (lower panel) for a system with small
perturbations, $L=50$, $x=0.26$ and $t_w=10^5$. $C_{tot}$ corresponds
to the autocorrelation function defined in the text. $C_{22}$ measures
the autocorrelation only in the subspace of sites with
$B(\vec{r},t_w)=2$.}
\end{center}
\end{figure}

In a very recent work in collaboration with
D. Stariolo~\cite{Copelli03} we have used the auto-correlation
functions (a common tool to study aging effects in e.g. glassy
systems) in order to analyze the similarities between the aging
effects in the multi-connected network and in glassy systems.  While
in glassy systems the aging effects results from the loss of
plasticity generated by frustration mechanisms, in the network these
effects are caused by multiple perturbations, since in order to
incorporate information the system has to adapt to (or satisfy) the
mechanisms of activation and suppression of the dynamics. The usual
procedure adopted to study glassy systems consists in taking a
``picture'' of the system at time $t_w$ and calculating the number of
sites that do not change during the following time steps. In this
sense the auto-correlation function is equal to
$C_{tot}(t,t_w)=1-hd(t,t_w)=1-HD(t,t_w)/N$, where $hd(t,t_w)$
[$HD(t,t_w)$] is the normalized (non-normalized) Hamming distance
between the configurations at times $t$ and $t_w$. According to the
results obtained, the system without any perturbation is driven to a
long period attractor after a long transient time ($10^4-10^5$ time
steps). All the biologically relevant effects are observed in the
transient time of the system. In the purely deterministic case, the
transient is simply the time it takes for the system to reach the
attractor ($\sim 10^4$ time steps).  When subjected to random small
perturbations, however, the very notion of a transient becomes fuzzy:
results in Ref.~\cite{Copelli03} show that the system approaches a
cycle, but is deflected from it by the small perturbations after $\sim
10^3-10^4$ time steps. Therefore, the small perturbations will cause
the system to change attractors from time to time due to their
cumulative effects. This is reflected in the decrease of the
auto-correlation functions, as can be seen Fig.~\ref{fig:correlation},
where the system leaves the cycle it had approached until $t_w=10^5$
time steps (these results are discussed in detail in
Ref.~\cite{Copelli03}). The analogue observed in glassy systems
corresponds to changes of the system to different minima of the energy
landscape during the relaxation time. Contrary to one's initial
intuition that the large perturbations would accelerate the
de-correlation process, the changes induced by large perturbations do
not lead to this effect. Curiously, the large perturbations alone
(which correspond to the immunizations) lead to a much weaker (slower)
de-correlation: this is due to the fact that they are produced always
at the same sites in shape space. Small perturbations can be more
easily absorbed by the system than the large ones since they involve
only local changes. The results from the autocorrelation functions
therefore indicate that the system first approaches a cycle (phase
space compression), but after some time the small perturbations
eventually deflect it from the trajectory to its ``natural''
attractor~\cite{Copelli03}.

\section{Results}
\label{results}

%%%%%%%%%%%%%%%%%%%%%%%%%%%%%%%%%%%%%%%%%%%%%%%%%%%%%
\begin{figure}[!bt]
\begin{center}
\includegraphics[width=0.7\textwidth,angle=-90]{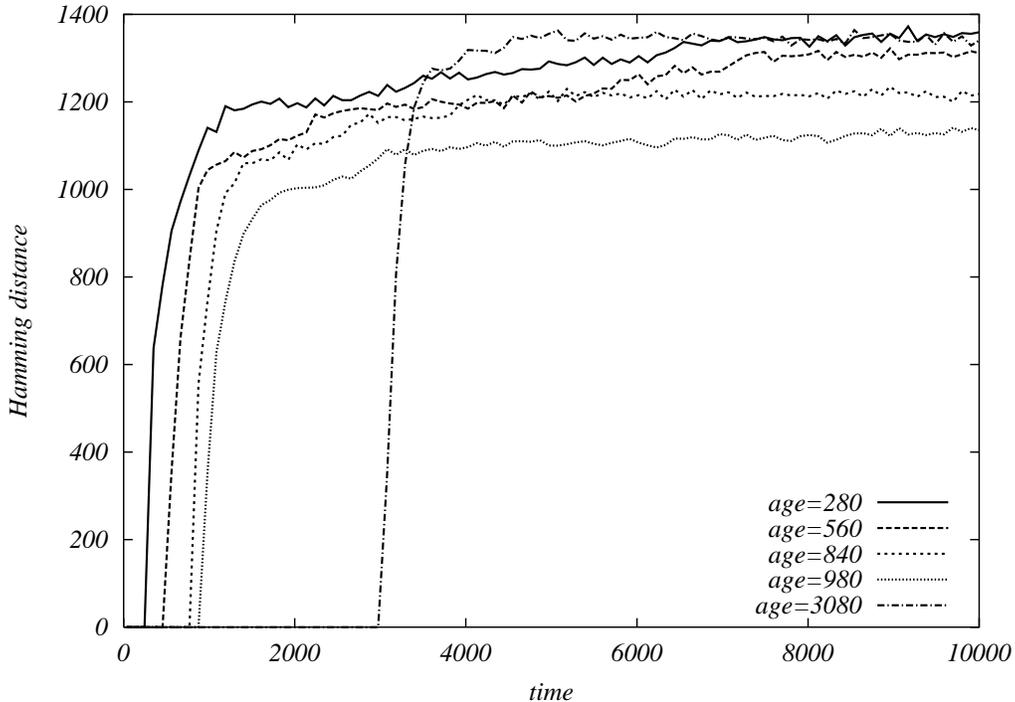}
\caption{\label{fig:HD}Mean Hamming distance vs. time for the complete
immunization protocol adopted in Ref.~\cite{zsb98} and for different
ages. Average over 10 runs. Standard deviations are not shown for the
sake of clarity.}
\end{center}
\end{figure}
%%%%%%%%%%%%%%%%%%%%%%%%%%%%%%%%%%%%%%%%%%%%%%%%%%%%%

This scenario provides a possible explanation for the results
regarding the spread of damage in the system, which managed to
reproduce the experimental data from immunization experiments with
mice~\cite{zsb98}. Fig.~\ref{fig:HD} shows the time evolution of the
Hamming distance between a system which undergoes only small
perturbations, and its copy, which undergoes the same small
perturbations {\em and\/} the large periodic immunizations. All curves
saturate after some time, just like the Elisa
measurements~\cite{zsb98}. More important, however, is the fact that
the value of the saturation depends on the {\em age\/} at which
immunization started. As the age increases, the saturation value
decreases, a result which was also observed in the mice and is
interpreted as a sign of loss of plasticity with
age~\cite{zsb98}. This might be related to the fact that the system is
in its transient, while trying to reach its long-period attractor. The
``older'' the system is, the less configurations available there are.

\begin{figure}[!tb]
\begin{center}
\includegraphics[width=0.7\textwidth,angle=-90]{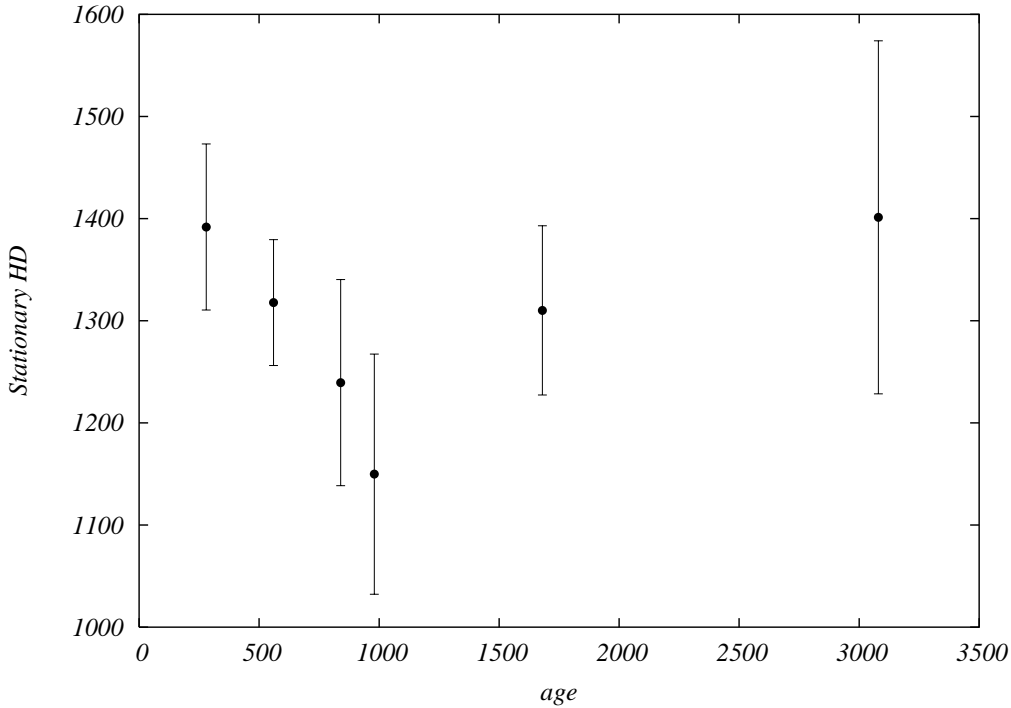}
\caption{\label{fig:stationary}Stationary Hamming distance vs. age at which
the immunizations start. Bars correspond to standard errors (average
over 10 runs).}
\end{center}
\end{figure}

If we now repeat the computer experiment of immunization (with small
and large perturbations) for ages greater than $\sim 10^3$ time steps,
we obtain a different result: for older systems the stationary Hamming
distance may increase, as shown in the rightmost curve of
Fig.~\ref{fig:HD}. We argue that the cumulative effects of the small
perturbations could be responsible for the change of behavior in the
rightmost curve of Fig.~\ref{fig:HD}: for sufficiently old systems,
the ``noise'' of the small perturbations become more important than
earlier in the transient, and the monotonicity of the stationary value
of the HD with the age at which immunization starts, no longer
holds. This result is summarized in Fig.~\ref{fig:stationary}. It is
interesting to point out that this breakdown of the refractoriness
induced by age, takes place on time scales which are comparable
(typically larger than) the life time of the mice.

\begin{figure}[!tb]
\begin{center}
\includegraphics[width=0.7\textwidth,angle=-90]{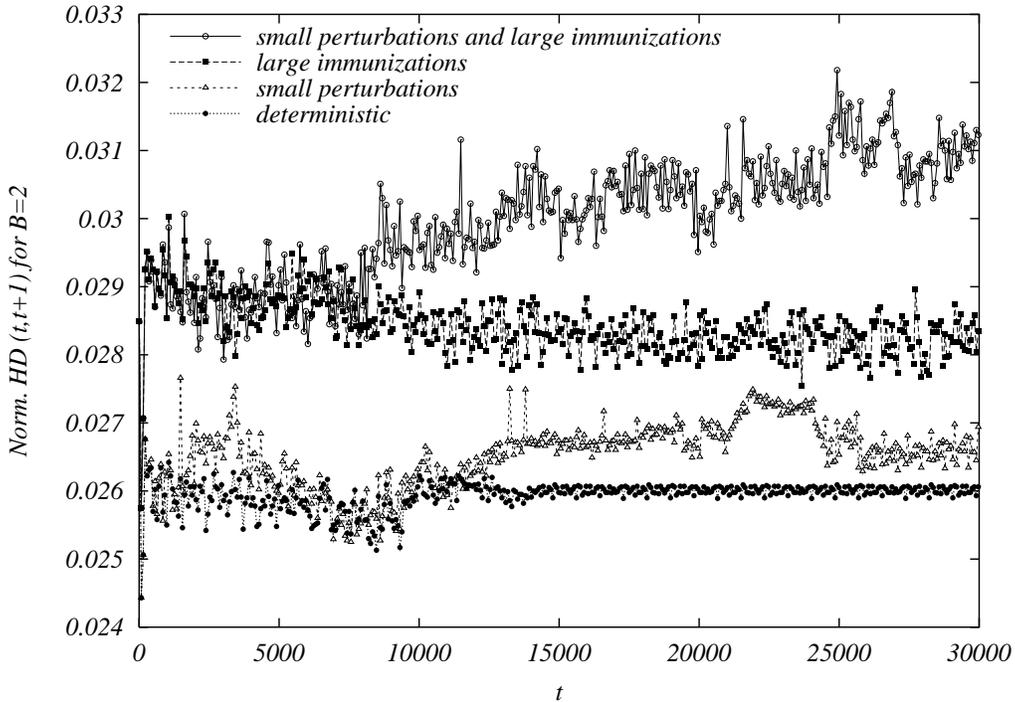}
\caption{\label{fig:twotimes}Normalized Hamming distance between configurations at times $t$ and
  $t+1$, taking into account only activated ($B=2$) sites.}
\end{center}
\end{figure}

The large perturbations, on the other hand, propagate to the entire
system and some of the active populations generated during the antigen
presentation will remain as part of the network. This mechanism
depends on the plasticity of the system or, in other words, will be
controlled by the characteristic cluster size of the
system~\cite{bzs01}. When we combine small and large perturbations,
the system is driven to a cycle that incorporates information about
the antigen, but will eventually be deviated from that trajectory as
soon as the cumulative effects of the small perturbations dominate the
dynamics~\cite{Copelli03}. In order to observe the {\em instantaneous
effects\/} of both types of perturbations, we have analyzed the
normalized Hamming distance between the same system at two consecutive
time steps, focusing only on the subspace of sites with
$B(\vec{r},t)=2$. This quantity is denoted by $HD_{22}(t,t+1)$ and
referred to simply as ``instantaneous Hamming distance''. It is
plotted in Fig.~\ref{fig:twotimes} as a function of time for different
situations.

\begin{figure}[!tb]
\begin{center}
\includegraphics[width=0.7\textwidth,angle=0]{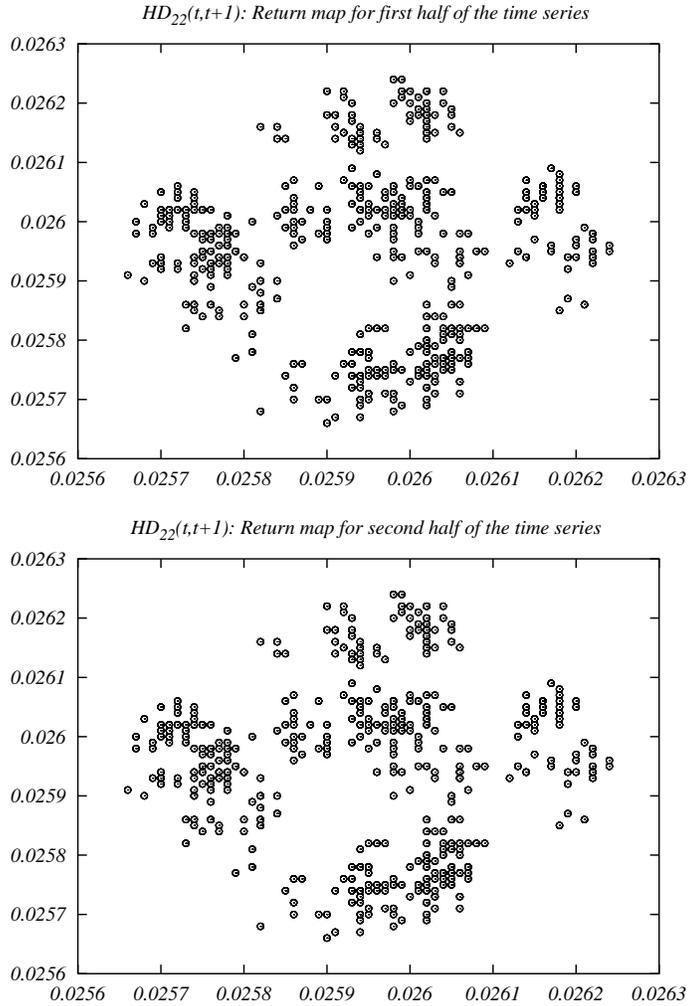}
\caption{\label{fig:return}Top panel: return map of $HD_{22}(t,t+1)$ from $t=20000$ to
$t=25000$.  Bottom panel: return map of $HD_{22}(t,t+1)$ from
$t=25001$ to $t=30000$. }
\end{center}
\end{figure}

We start the analysis of Fig.~\ref{fig:twotimes} by noticing the
behavior for the purely deterministic case (lower curve): the plot
indicates that typically $2.6\%$ of the active network (sites with
$B(\vec{r},t)=2$) changes at each time step. The end of the transient
is clearly seen at $t\sim 15000$ time steps (for this particular
realization): the curve then becomes periodic, with a very long
period. Note that the periodicity of the dynamics can be confirmed by
checking that the $C_{tot}(t_w,t)=1$ for some $t>t_w$ and for
sufficiently large $t_w$ (see Fig.~\ref{fig:correlation} and
Ref.~\cite{Copelli03}). To illustrate the periodicity of
$HD_{22}(t,t+1)$, we plot in Fig.~\ref{fig:return} the return map of
the time series from $t=20000$ to $t=25000$ time steps (upper
plot). The extremely long period might lead one to suspect, based on
the return map, that the system might be chaotic. This suspicion is
ruled out when one observes the return map from $t=25000$ to $t=30000$
time steps (lower plot). We challenge the reader to find a discrepancy
between the two return maps.

Returning to Fig.~\ref{fig:twotimes}, notice that the size of the
changes induced by the small perturbations changes near the end of the
transient time. While the perturbed system no longer attains the
cycle, it can still feel the effects of the attractor: when close to
the cycle, the small perturbations induce larger changes in the active
network at each time step. If subjected only to the large
immunizations, the instantaneous Hamming distance take typically
larger values. This might suggest a contradiction with the results in
Ref.~\cite{Copelli03}, according to which the system only with large
immunizations de-correlates slower. But the contradiction is only
apparent: $HD(t_w,t)$ measures a {\it long term effect\/}, which is
dominated by the small perturbations. Fig.~\ref{fig:twotimes}, on the
other hand, is a measure of the instantaneous change occurring in the
system, {\it i.e.\/} a {\it short time effect\/}. Finally, one can
combine both types of perturbation, the result being even larger
instantaneous changes.

%%%%%%%%%%%%%%%%%%%%%%%%%%%%%%%%%%%%%%%%%
%%%%%%%%%%%%%%%%%%%%%%%%%%%%%%%%%%%%%%%%%

Another aspect of Fig.~\ref{fig:twotimes} that should be noted is an
interesting interplay between the perturbations. In what follows, we
refer to the typical values of the four curves in
Fig.~\ref{fig:twotimes} (for large $t$), from bottom to top, as $d$
(``deterministic''), $s$ (``small''), $l$ (``large'') and $ls$
(``large and small'') --- therefore $d<s<l<ls$, as discussed
above. The fact that $ls-s>l-d$ indicates that the effect of the large
immunizations (i.e. the instantaneous change it induces) is boosted by
the presence of the small perturbations. The reverse is obviously also
true: the previous inequality implies $ls-l>s-d$, indicating that the
small perturbations induce larger changes in the network if the large
immunizations are present. This signals a cooperative interaction
between two kinds of perturbations which are significantly different
in nature, leading to the necessary complexity of behavior which allow
the model to reproduce experiments performed with real immune systems.

\section{Concluding Remarks}
\label{conclusao}

In this paper we have reviewed the main results from the biological
and physical point of view of a cellular automata model introduced by
SW in 1992. Since it incorporates the main features of the immune
network theory and models the functioning of a multi-connected
network, it emerged as a candidate to reproduce the basic dynamics of
what the real immune network could be. The model was then used to
model real experiments performed with mice about immunization and
aging. The immunization experiments are simulated by producing small
and large perturbations or damages. In this paper we complement the
results of a very recent work~\cite{Copelli03} with the study of the
role of each kind of perturbation. The results obtained in the
previous work, using auto-correlation functions, apparently suggested
that the small perturbations would be more important than the large
ones from the dynamical point of view, since their cumulative effects
could change the attractor of the dynamics after a few thousand time
steps. However this study was focusing in the long term behavior of
the system.  Here we have studied its short term behavior and shown
that both types of perturbation are important: it is their {\em
cooperative effects\/} that generate the necessary complexity which
allowed to reproduce experiments performed with real immune systems.

Despite the understanding of the dynamics we have achieved up to this
point, there are still some points that should be investigated: what
happens when changing the frequency of presentation of the small
perturbations for a fixed value of the large damages, and how these
results will change when varying the large damage size? This study
(already in progress) will draw a scenario that would allow us to try
to apply this model to reproduce other experimental results
concerning, for instance, cross-reactivity mechanisms observed in
immunization protocols, and tolerance.

\bibliography{copelli}

\end{document}